 \documentclass[preprint2]{aastex}

\shorttitle{GRP-VLBI}
\shortauthors{Takefuji et al.}

\usepackage{natbib}
\usepackage{amsmath} % for aligh
\usepackage{lscape} % table landscape
\usepackage{url} 
\begin{document}

%% LaTeX will automatically break titles if they run longer than
%% one line. However, you may use \\ to force a line break if
%% you desire.

\title{Very Long Baseline Interferometry Experiment \\ on Giant Radio Pulses of Crab Pulsar \\ toward Fast Radio Burst Detection}

%% Use \author, \affil, and the \and command to format
%% author and affiliation information.
%% Note that \email has replaced the old \authoremail command
%% from AASTeX v4.0. You can use \email to mark an email address
%% anywhere in the paper, not just in the front matter.
%% As in the title, use \\ to force line breaks.

\author{K. Takefuji\altaffilmark{1}, T. Terasawa\altaffilmark{2}, T. Kondo\altaffilmark{1}, R. Mikami\altaffilmark{2}, H. Takeuchi\altaffilmark{3}, H. Misawa\altaffilmark{4}, F. Tsuchiya\altaffilmark{4}, H. Kita\altaffilmark{4} and M. Sekido\altaffilmark{1}}

\altaffiltext{1}{National Institute of Information and Communications
Technology, 893-1 Hirai, Kashima, Ibaraki 314-8501, Japan.} 
\altaffiltext{2}{ Institute for Cosmic Ray Research, The University of Tokyo, Kashiwa, Chiba 277-8582, Japan}

\altaffiltext{3}{ Institute of Space and Astronautical Science, Japan Aerospace Exploration Agency, Sagamihara, Kanagawa 252-5210, Japan}

\altaffiltext{4}{ Planetary Plasma and Atmospheric Research Center, Tohoku University, Sendai, Miyagi 980-8578, Japan}

\email{takefuji@nict.go.jp}

%% Notice that each of these authors has alternate affiliations, which
%% are identified by the \altaffilmark after each name.  Specify alternate
%% affiliation information with \altaffiltext, with one command per each
%% affiliation.

%% Mark off your abstract in the ``abstract'' environment. In the manuscript
%% style, abstract will output a Received/Accepted line after the
%% title and affiliation information. No date will appear since the author
%% does not have this information. The dates will be filled in by the
%% editorial office after submission.

%%    ABSTRACT    %%
\begin{abstract}
 We report on a very long baseline interferometry (VLBI) experiment on giant radio pulses (GPs) from the Crab pulsar in the radio 1.4 to 1.7 GHz range to demonstrate a VLBI technique for searching for fast radio bursts (FRBs). We carried out the experiment on 26 July 2014 using the Kashima 34 m and Usuda 64 m radio telescopes of the Japanese VLBI Network (JVN) with a baseline of about 200 km. During the approximately 1 h observation, we could detect 35 GPs by high-time-resolution VLBI. Moreover, we determined the dispersion measure (DM) to be $56.7585 \pm 0.0025$ on the basis of the mean DM of the 35 GPs detected by VLBI. We confirmed that the sensitivity of a detection of GPs using our technique is superior to that of a single-dish mode detection using the same telescope. 
   
\end{abstract}

%% Keywords should appear after the \end{abstract} command. The uncommented
%% example has been keyed in ApJ style. See the instructions to authors
%% for the journal to which you are submitting your paper to determine
%% what keyword punctuation is appropriate.


%% From the front matter, we move on to the body of the paper.
%% In the first two sections, notice the use of the natbib \citep
%% and \citet commands to identify citations.  The citations are
%% tied to the reference list via symbolic KEYs. The KEY corresponds
%% to the KEY in the \bibitem in the reference list below. We have
%% chosen the first three characters of the first author's name plus
%% the last two numeral of the year of publication as our KEY for
%% each reference.

%% Authors who wish to have the most important objects in their paper
%% linked in the electronic edition to a data center may do so by tagging
%% their objects with \objectname{} or \object{}.  Each macro takes the
%% object name as its required argument. The optional, square-bracket 
%% argument should be used in cases where the data center identification
%% differs from what is to be printed in the paper.  The text appearing 
%% in curly braces is what will appear in print in the published paper. 
%% If the object name is recognized by the data centers, it will be linked
%% in the electronic edition to the object data available at the data centers  
%%
%% Note that for sources with brackets in their names, e.g. [WEG2004] 14h-090,
%% the brackets must be escaped with backslashes when used in the first
%% square-bracket argument, for instance, \object[\[WEG2004\] 14h-090]{90}).
%%  Otherwise, LaTeX will issue an error. 

%%%%%%%%%%%%%%%%%%%%%%%%%%%%%%%%%%%%%%%%%%%%%%%%%%%%%%%%%%%%%%%%%%%%%%%%%
%%     INTRODUCTION      %%
\section{Introduction}  
\cite{2007Sci...318..777L} reported an interesting burst phenomenon, the so-called fast radio bursts (FRBs), which comprised a single impulse of 30 Jy with a frequency-dependent  dispersion having  a dispersion measure (DM) of 375 pc cm${}^\text{-3}$, in wide-field pulsar surveys using the 13-beam, 1.4 GHz receiver at the Parkes radio telescope.  This large DM with a Galactic latitude of $-42^\circ$ indicates that the pulse has a cosmological origin at $z \sim 0.3$. After the discovery of FRBs, some FRBs were detected with the Parkes radio telescope (\citealt{2007Sci...318..777L}, \citealt{2013Sci...341...53T}, \citealt{2014ApJ...792...19B}) and  FRB110523 detected with the Green Bank telescope revealed Faraday rotation (\citealt{2015Natur.528..523M}), and FRB121102 detected with the Arecibo radio telescope (\citealt{2014ApJ...790..101S}). \cite{2016Natur.531..202S} reported the detection of ten additional bursts, which have dispersion measures and sky positions consistent with the original burst of FRB121102,  using the Arecibo telescope. Although most reported FRBs were discovered later in archival data,
	FRB150418 was recently detected in real-time by the Parkes 64 m radio telescope, which triggered follow-up observations at wavelengths of 5.5 GHz and 7.5 GHz by the Australia Telescope Compact Array (ATCA) 2 h after the event and optical observation by Suprime-Cam on the 8.2 m Subaru telescope within the $\sim$ 1 arcsec positional uncertainty derived from the ATCA image in the following two days (\citealt{2016Natur.530..453K}). \cite{2016Natur.530..453K} identified an elliptical galaxy as the host galaxy of FRB150418,  whose redshift was measured to be $z = 0.492 \pm 0.008$ with a fading radio transient lasting $\sim$ 6 days after the event. Exciting FRB astronomy is now underway. However, in order to understand the property of FRBs astrophysically, a large number of FRB events, multi-frequency counterpart identification, and host identification are required.
	
Astronomical and geodetic very long baseline interferometry (VLBI) (e.g., Very Long Baseline Array (VLBA), European VLBI Network (EVN), The Australian Long Baseline Array (LBA), and East-Asia VLBI Network (EAVN)) is routinely performed throughout the year while observing almost the entire sky. Thus, if we detect FRBs by VLBI observations, the number of detections could be increased and the detection by VLBI could provide essential scientific information in terms of FRB localization. However, these VLBI observations have not been dedicated to finding only FRBs. We should perform these observations as a by-product without disturbing the main observations.  \cite{2011ApJ...735...97W} carried out an experiment called V-FASTR to perform a blind search for fast transient radio signals using VLBA. Since they searched for FRBs using several single dishes operated in parallel, it was not a ''true" VLBI experiment. If a survey of FRBs  is carried out by VLBI, the sensitivity must be higher than that of single-dish detection methods using same telescopes. 
In order to explore the possibilities of FRB detection by VLBI, we carried out a true VLBI experiment by observing the Crab pulsar, which often produces very large pulses, so-called giant radio pulses (hereafter GPs). Currently over 2000 pulsars have been found, although only a handful of them have been found to emit GPs (\citealt{2006ChJAS...6b..41K}). 
Individual GPs from the Crab pulsar reach brightness temperatures of at least $10^{32}$ K (\citealt{2004ApJ...612..375C}), which do not resolve the narrowest pulses. GPs are known to reach $10^{37}$ K in nanosecond-resolution observations (\citealt{2003Natur.422..141H}). Since the GPs from the Crab pulsar are dispersed for a DM of about 56.8 pc cm${}^\text{-3}$, a de-dispersion process is required to reconstruct a sharp pulse. Thus,
since the GPs from the Crab pulsar have a short time scale and a dispersed pulse, they have similar characteristics to FRBs.
\cite{2015arXiv150500987R} reported a preliminary VLBI result for a GP from the Crab pulsar with Radioastron at space-ground baselines, and the correlation for space-ground baselines was found in four sessions at 18 cm. They estimated several parameters to obtain primary information about scattering effects in the interstellar medium using a changing correlation amplitude that depended on the baseline projection. 
In the field of transient radio phenomena, it is necessary to announce a detection as soon as possible. Normally, we do not perform searches of baseband data because the recording data rate is too high to process.  A feasible procedure for searching for transient impulses at a high speed is required. As reported in this paper, we found correlations of GPs from the Crab pulsar by VLBI. The correlations appeared as sharp peaks in the cross-spectrum, which we will use as a trigger for an alert signal without disturbing the correlation process.
  Here, we report the VLBI observation of the Crab pulsar and determine the DM by newly developed high-time-resolution VLBI.  
%Our motivation is to detect a correlation of GP from Crab pulsar by newly developed high-time-resolution VLBI toward FRB detection. 

%%%%%%%%%%%%%%%%%%%%%%%%%%%%%%%%%%%%%%%%%%%%%%%%%%%%%%%%%%%%%%%%%%%%%%%%%
%%     Observations     %%
\section{Observations }
We carried out the 1 h VLBI experiment from 23h UTC on 26 July 2014 using the Kashima 34 m and Usuda 64 m radio telescopes, the latter having the largest antenna in Japan. The baseline length between the two antennas is about 200 km. We used the L-band and the right-hand circular polarization of both radio telescopes. The L-band receiver of the Kashima 34 m radio telescope has cooled band-pass filters, whose pass-bands have frequency ranges of 1405 to 1440 MHz and 1600 to 1720 MHz, installed before the low-noise amplifier to suppress external interference.
The system equivalent flux density (SEFD) of the Kashima 34 m and Usuda 64 m radio telescopes are typically 400 and 100 Jy, respectively.

The downconverted received signal of the L-band receiver was transferred to an observation room. Then, the intermediate signal with a frequency below 512 MHz was digitally baseband-converted (DBBC) and four-bit-recorded at a sampling rate of 64 MHz using an ADS3000+ digital sampler, which has 16 DBBC channels (\citealt{2010ivs..conf..378T}). The 32 MHz frequency ranges of the seven recording channels were tuned for the bandwidths of 1400 to 1432 MHz, 1411 to 1443 MHz, 1570 to 1602 MHz, 1602 to 1634 MHz, 1634 to 1666 MHz, 1666 to 1698 MHz, and 1698 to 1730 MHz. However, with the seven channels fully covered with the above pass-bands of the L-band receiver, we did not use the other DBBC channels. The basebands of the seven channels were digitally recorded at the Nyquist rate with a 64 MHz sampling speed.
Before the Crab pulsar observation, we observed quasar 0552+398 for 30 s from 23:04 on 26 July 2014 to determine the  clock offset between the seven channels.
  %%%%%%%%%%%%%%%%%%%%%%%%%%%%%%%%%%%%%%%%%%%%%%%%%%%%%%%%%%%%%%%%%%%%%%%%%
%%     Resultss     %%
\section{Results }
Firstly, we performed a correlation using a GICO3 software correlator (\citealt{2012ivs..conf...91O})	 to measure the clock offset, and thus obtained the correlation of quasar 0552+398 between Kashima 34 m and Usuda 64 m. The signal-to-noise ratio (SNR) of each channel became over 150 with  30 s integration. The measured clock offsets for the second to the seventh channels relative to the delay in the first channel were 2.078 ns, 18.751 ns, 22.249 ns, 23.007 ns, 21.987 ns, and 21.833 ns, respectively. 
  
\begin{figure}[hbt]
  \includegraphics[width=\linewidth]{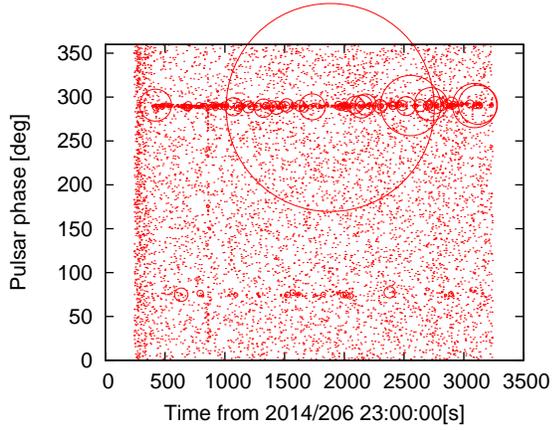}
  \caption{Graph of Crab pulsar phase plotted against time during  1 h observation showing main pulse at 290 deg and interpulse at 75 deg. The sizes of the circles represent the SNR for the GPs. We used the first channel of Kashima 34 m with the frequency range of 1400 to 1432 MHz. We performed the dedispersion process using a DM of 56.78 pc cm${}^\text{-3}$ with 1 $\mu s$ integration. When we adjusted the pulsar cycle to 33.6963 ms, the series of GPs aligned with the pulsar phase. The reference point of  the pulsar phase is 00:00 on 26 July 2014. }
  \label{fig:single}
\end{figure}

  Next, we performed the de-dispersion only for the first channel of Kashima 34 m with the frequency range of 1400 to 1432 MHz to find GP candidates. Figure \ref{fig:single} shows the obtained Crab pulsar phase plotted against time during the 1 h observation. We used a DM of 56.78 pc cm${}^\text{-3}$ referenced by the Jodrell Bank monthly ephemeris of the Crab pulsar\footnote{\url{http://www.jb.man.ac.uk/pulsar/crab.html}} (\citealt{1993MNRAS.265.1003L}) and 1 $\mu s$ integration. As a result, we found 71 GPs with an SNR above 18. The strongest GP during the observation had an SNR of 504 and was observed at 23:31:22.11 on 26 July 2014. Next, we will perform the VLBI correlation on the baseband data of the 71 GPs found by single-dish observation.  
  
  Now, we will explain the procedure of VLBI correlation for GPs with a GICO3 software correlator. When we process normal VLBI data with digital FX-type spectrometers, Fourier transformation and multiplication are applied to the data to obtain a power spectrum of the signal. However, GPs and FRBs, which have swept-frequency characteristics, become  smeared pulses after performing the  integration. To avoid smeared pulses, we skipped the integration process in GICO3 to realize a high time resolution for processing a GP signal. We performed this high-time-resolution process on the strongest GP from the seventh digitized baseband data as an example. 
  
\begin{figure}[hbt]
  \includegraphics[width=\linewidth]{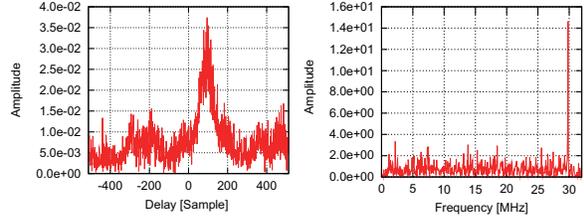}
  \caption{VLBI result for the strongest GP from the Crab pulsar on 26 July 2014. The right graph shows the strong GP cross-spectrum in the frequency domain at 1728 MHz in the case of 16 $\mu$s integration at 23:31:22.118672. The left graph shows the weak and broad GP correlation in the time domain in the case of 400 $\mu$s integration at 23:31:22.1212. }
  \label{fig:fringe}
\end{figure}

   Figure \ref{fig:fringe} shows the correlation of the strongest GP in both the time and frequency domains with 1024 Fourier lengths per 32 MHz bandwidth (62.5 kHz resolution). A sharp peak with SNR of 35 appears in the right-side cross-spectrum in Figure \ref{fig:fringe} at 23:31:22.118672 from a single Fourier period of 16 $\mu$s (1024 Fourier lengths in 64 MHz sampling). In contrast to the frequency domain, although we observed no correlation from a single Fourier period in the time domain, we  needed more time in the case of 400 $\mu$s integration to accumulate 25 Fourier bins to find the weak and broad correlation on the left of Figure \ref{fig:fringe} at 23:31:22.1212. Such a weak and broadened signal with SNR of 4.8 in the time domain would be difficult and ineffective to search for. However, sharp peaks that appeared in the Fourier domain could be used as a trigger for an alert signal without disturbing the correlation process.

\begin{figure}[hbt]
  \includegraphics[width=\linewidth]{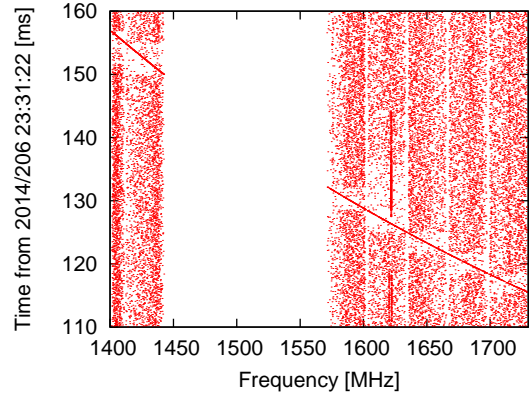}
  \caption{ Swept-frequency curve of the strongest GP from the Crab pulsar detected from the correlation between Kashima 34 m and Usuda 64 m at 23:31:22 on 26 July 2014, shown as  stacked extracted peaks from every 16 $\mu$s cross-spectra. An artificial signal can be seen at 1620 MHz. However, an artificial signal with a specific frequency can be clearly distinguished and easily removed. We estimated the DM to be  56.7620 $\pm$ 0.0003 from the curve. }
  \label{fig:grp-bws}
\end{figure}

  For the strongest GP, we next stacked all Fourier bins in which the GP existed  for about 40 ms in all seven channels. Although the amount of cross-spectrum data became huge (about 72 MB in only 40 ms), we only extracted the peak frequency and its amplitude to reduce the amount of data. As described previously, the maximum time difference between the seven channels was 23 ns. Since this is smaller than the single Fourier length of 16 $\mu$s, we ignored the time differences. Then, we stacked the extracted peaks from every Fourier bin as shown in Figure \ref{fig:grp-bws}. In total, 22,000 (50 ms / 16 $\mu$s $\times$ 7 bands) data  points were included in the figure. A GP curve originating from the dispersed delay caused by the interstellar medium and an artificial signal at 1620 MHz possibly from a Japanese communication satellite were observed. However, an artificial signal with a specific frequency can be clearly distinguished and easily removed. 
% Whether signals has an astronomical or terrestrial origin can be distinguished them by VLBI using two  separately located antennas on Earth. Since such antennas are affected by different Doppler shifts owing to Earth's rotation, signals with a terrestrial origin disappear when the Doppler shifts are synchronized in the correlation process to data from the two antennas when they are oriented toward a point in the sky. 
  
  Next, we estimated the DM from the dispersed curve in Figure \ref{fig:grp-bws}.
   The observation equation with respect to the arrival time of the GP, $\tau(f)$, is,
 \begin{equation}
   \tau(f) = \frac{e^2}{2\pi m_e c}\frac{DM}{ f^{2} }+t_0, \label{eqn:tau}
 \end{equation}
where $e$ is the electron charge, $m_e$ is the electron mass, $c$ is the speed of light, and $t_0$ is the arrival time at the highest frequency. Although it was easy to estimate the unknown parameters $DM$ and $t_0$ from Figure \ref{fig:grp-bws} by the least-squares method (LSM) using a standard gnuplot, the other weaker GPs could not be fitted using the gnuplot. Thus, we wrote an automated fitting program with the C++ template library Eigen\footnote{\url{http://eigen.tuxfamily.org}} for linear algebra. The functions of our LSM program are as follows:
  \begin{itemize}
  \item Calculating and removing the interference in specific frequency migration by stacking at every frequency resolution automatically.
  \item Iterative fitting by removing noise outside 3$\sigma$ from the GP curve and narrowing the time (initially 50 ms) by 99\% in slow steps to prevent a local minimum and calculating the chi-square value.
  \item Estimating $DM$ and $t_0$ when the chi-square value becomes within $1\pm0.01$.
\end{itemize}

\begin{figure}[hbt]
  \includegraphics[width=\linewidth]{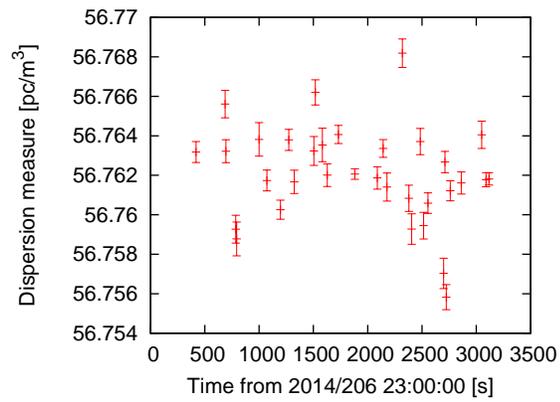}
  \caption{Estimated DM from 35 GPs observed by VLBI. The mean DM is $ 56.7620 \pm 0.0025$. The final DM is $56.7585 \pm 0.0025$ considering the Doppler effects of the Earth's revolution.}
  \label{fig:dm-determination}
\end{figure}

We differentiate equation (\ref{eqn:tau}) with respect to the frequency $f$. $\tau(f)$ has an uncertain value, which we used as the weight for the chi-square calculation, with regard to the observation frequency,  
 \begin{eqnarray}
   \frac{d\tau(f)}{df} &=& -\frac{e^2}{\pi m_e c}\frac{DM}{ f^{3} } \\
 &=& -2\frac{\tau(f)}{f}.
 \end{eqnarray}
 %When we observe a pulsar at 1600 MHz, DM is 56.7 and $df$ is 62.5 kHz (=64 MHz sampling / 1024 Fourier point), the $d\tau$ becomes 7.18 us uncertainly.
  Finally, we performed the previous process, which includes a correlation, stacking the peaks, and fitting, routinely for the 71 GPs already found by single-dish observation and obtained a DM of 56.754 to 56.770 from the 35 GPs detected by VLBI. The mean value of the DM was estimated to be $ 56.7620 \pm 0.0025$, as shown in Figure \ref{fig:dm-determination}. The final DM considering the Doppler effects of the Earth's revolution is $56.7585 \pm 0.0025$. Our DM estimated by VLBI is 2$\sigma$ close to 56.7632, the value obtained on 14 July 2014  from  the Jodrell Bank Crab pulsar monthly ephemeris. This small difference is caused by the fact that the DM is observed  to change with time for many pulsars because of solar wind (e.g., \citealt{2007MNRAS.378..493Y}), and the movements of the pulsar, the Earth, and the interstellar medium (e.g., \citealt{2013MNRAS.429.2161K}). \cite{2007MNRAS.378..493Y} reported that a solar wind contributes $\sim$ 100 ns at 1400 MHz for sources within $60^{\circ} $ of the Sun and $\sim$ 1 $\mu$s for sources within $7^{\circ}$ of the Sun. Since we observed the Crab pulsar within about $45^{\circ}$ of the Sun, we assume that the solar wind contributes approximately 250 ns by interpolating the above relationship and taking the log of the times, which corresponds to a DM variation on the order of $10^{-3}$.     
  %%%%%%%%%%%%%%%%%%%%%%%%%%%%%%%%%%%%%%%%%%%%%%%%%%%%%%%%%%%%%%%%%%%%%%%%%
%%    Discussion   %%
\section{Discussion}
   Since the numbers of GPs detected by VLBI observation 35 was less than the 71 GPs (SNR $> 18$) extracted from only a de-dispersed single 32 MHz channel, the VLBI technique that we applied is disadvantageous compared with the single-dish observation. This is because we used only one peak frequency in the 16 $\mu$s cross-spectrum bin, which had a total of 1024 data in this case.
   For the sake of comparison between single-dish and VLBI observation, we performed the same VLBI process for the 71 GPs obtained using a single dish (Kashima 34 m or Usuda 64 m). In the VLBI, we performed our processing on  the cross-spectrum.  In contrast, we performed the processing on the power spectra of both Kashima and Usuda for the single-dish observations.  We counted the GP residuals after fitting from 1400 to 1730 MHz as the comparison method. The result is shown in Table \ref{tbl:count}. For example, the strongest GP at 23:31:22 in line 17 has 1085 GP residuals for the case of VLBI. However, the two single-dish observations found 842 and 935 GP residuals for Kashima and Usuda, respectively. Thus, the results obtained by VLBI were superior to those obtained by the single-dish observation, and were on average 1.6 times higher.
Since the VLBI result in the first row in the table existed,  some pointing error might have occured in the case of Usuda 64 m. 
This VLBI result allows us to distinguish signals from the sky from local interference. For example, some artificial impulses are known to be formed as swept-frequency-dependent signals, which are very similar to FRBs (\citealt{2011ApJ...727...18B}, \citealt{2015MNRAS.451.3933P}). Since these locally emitted signals are not subjected to interference  over a 100 km baseline, we assume that these interferences could be easily neglected by this VLBI technique. 
In brief, the sensitivity of VLBI is given by
   \begin{equation}
  SNR = \frac{F}{\sqrt{SEFD_1}\sqrt{SEFD_2}}\sqrt{2BT}, \label{eqn:min}
\end{equation}
 where $F$ is the flux density, $SEFD_1$ and $SEFD_2$ are the SEFDs of the two antennas, $2B$ is the Nyquist sampling speed, and $T$ is the integration time. In our case, the sampling speed $2B$ was 64 MHz and $T$ was 16 $\mu$s. If we determine SNR as a threshold for searching for the peak to be 18, the sensitivity from equation (\ref{eqn:min}) becomes 79.5 Jy as a snapshot, such as right figure in Figure \ref{fig:fringe}. If we integrate the whole 150 MHz on seven channels in our VLBI, the sensitivity becomes 30.1 Jy. Thus, we will perform the de-dispersion on the cross spectrum to acquire better sensitivity as a next study. 
 
   Toward the detection of FRBs by VLBI, we propose that all institutes and universities with radio telescopes should observe the same area of the sky and record data while synchronizing the frequency bandwidth during vacant time.  If FRBs occur and such data are obtained, we can apply our VLBI technique to detect them. This will increase the detection of  FRBs and we can determine the exact localization of the FRBs. Moreover, we normally only search for peaks in the time domain for geodetic and astronomical VLBIs, especially for continuum sources. On the basis of our results, we propose that peaks should also be searched for in the frequency domain. Normally we perform correlation by  Fourier transforming the data of two stations and multiplying the two spectra to form a cross spectrum. The largest load in the correlation process is the fast Fourier transform (FFT). After every FFT, a peak search in the frequency domain is performed. If we define $N$ as the Fourier length, the processing costs of a simple radix-2 FFT and the linear search are $Nlog_{2}N$ and $N$, respectively. In our case, we used 1024 Fourier lengths for a 32 MHz bandwidth (62.5 kHz resolution). This is expected to increase the processing time by 10\% as a result of adding a linear search for a peak in the correlation. If we perform the correlation  at a rather coarse frequency resolution of 1 MHz as  general VLBI processing, the processing time is increased by 20\%. If we change the general VLBI processing ($N$=32) to high-time-resolution VLBI ($N$=1024), the processing time is increased by 220\%. However, the increase in processing time for the high-time-resolution VLBI will be cancelled out by the recent progress in computers. 
   
%%    Summary   %%
\section{Summary}   
 We carried out a high-time-resolution VLBI experiment on GPs from the Crab pulsar using the Kashima 34 m and Usuda 64 m radio telescopes in the 1.4 to 1.7 GHz range. We obtained a sharp correlation in the case of 16 $\mu$s integration in the frequency domain after determining the clock offset between two antennas. We also obtained the DM of the GPs from a curve obtained by stacking peaks from every cross-spectrum. Then, we estimated the DM to be $ 56.7585 \pm 0.0025$ from the 35 detected GPs. The technique we applied has the potential to detect FRBs.   
% \section*{Acknowledgements}
 
 %%%%%%%%%%%%%%%%%%%%%%%%%%%%%%%%%%%%%%%%%%%%%%%%%%%%%%%%%%%%%%%%%%%%%%%%%
%%     Bibliography     %%

%\bibliography{ms.bib} % use bibtex
%\bibliographystyle{chicago}
%\bibliographystyle{plain}

%%%%%%%%%%%%%%%%%%%%%%%%%%%%%%%%%%%%%%%%%%%%%%%%%%%%%%%%%%%%%%%%%%%%%%%%%
%%     Tables     %%

%\renewcommand{\baselinestretch}{1} % for english revise
\clearpage
\begin{table}[bp]
\begin{center}
\begin{tabular}{cccccccc} \hline
No. & Epoch           &Type& SNR   & VLBI   & KASHIMA & USUDA & Increasing rate\\
   &                 &    &       &[count] &[count] & [count] & \\ \hline
 1 & 23:06:59.509753 & MP & 82.57 &  250   & 599 &     & (0.42)\\
 2 & 23:11:27.462469 & MP & 22.14 &  156   &     & 94  & (1.66)\\ %&
 3 & 23:11:34.538977 & MP & 22.67 &  226   & 117 & 152 & 1.69\\
 4 & 23:13:05.148274 & MP & 20.66 &  144   &     & 97  & (1.48)\\
 5 & 23:13:13.417519 & IP & 21.34 &  84    &     & 65  & (1.29)\\ %&
 6 & 23:16:40.534982 & MP & 25.44 &  94    &     &     & \\
 7 & 23:17:52.105830 & MP & 49.20 &  290   & 181 & 204 & 1.51\\
 8 & 23:18:58.622391 & MP & 21.47 &  78    &     &     &\\ %%
 9 & 23:19:56.478841 & MP & 34.37 &  371   & 197 & 248 & 1.68\\
10 & 23:21:13.643400 & MP & 22.77 &  282   & 153 & 150 & 1.86\\
11 & 23:22:04.490966 & MP & 50.18 &  281   & 152 & 195 & 1.63\\
12 & 23:25:03.081545 & MP & 28.91 &  124   &     &     &\\ %%
13 & 23:25:18.885214 & MP & 38.27 &  191   & 104 & 132 & 1.63\\
14 & 23:26:22.975541 & MP & 20.29 &  86    & 42  &     & (2.05)\\
15 & 23:27:08.802507 & MP & 32.57 &  240   & 126 & 127 & 1.90\\
16 & 23:28:50.632612 & MP & 68.07 &  386   & 245 & 253 & 1.55\\
17 & 23:31:22.131100 & MP & 504.1 &  1085  & 842 & 935 & 1.22\\
18 & 23:33:19.394396 & MP & 22.90 &  111   &     &     &\\
19 & 23:34:48.352736 & MP & 47.91 &  224   & 138 & 119 & 1.75\\
20 & 23:35:42.670998 & MP & 60.83 &  330   & 158 & 221 & 1.77\\
21 & 23:36:18.995826 & MP & 56.45 &  138   &     & 87  & (1.59)\\ 
22 & 23:38:39.542937 & MP & 29.91 &  146   &     &     &\\
%ND& 23:38:19.089261 & MP & 38.28 &        &     &   \\
23 & 23:39:39.704649 & IP & 33.00 &  240   & 125 & 200 & 1.52\\
24 & 23:40:05.940311 & MP & 25.08 &  106   &     & 74  & (1.43)\\
25 & 23:41:25.429876 & MP & 30.09 &  152   & 100 & 115 & 1.42\\
26 & 23:41:55.992362 & MP & 51.40 &  166   & 73  & 113 & 1.83\\
27 & 23:42:35.181207 & MP & 150.5 &  284   &     & 177 & (1.60)\\
28 & 23:44:59.772031 & MP & 27.71 &  116   &     &     &\\
29 & 23:45:12.677827 & MP & 22.69 &  256   & 129 & 183 & 1.67\\
30 & 23:45:24.437880 & MP & 85.05 &  190   & 110 & 126 & 1.61\\
31 & 23:46:01.975449 & MP & 47.22 &  289   & 157 & 188 & 1.68\\
%ND& 23:46:54.036312 & MP & 35.50 &        &     &  \\
%ND& 23:47:34.033695 & MP & 21.15 &        &     &  \\
32 & 23:47:42.626430 & MP & 25.84 &  268   &     & 204 & (1.31)\\
33 & 23:50:50.921363 & MP & 24.67 &  194   & 92  & 115 & 1.89\\
34 & 23:51:30.716503 & MP & 112.1 &  679   & 436 & 549 & 1.39\\
35 & 23:51:58.212767 & MP & 99.38 &  860   & 556 & 666 & 1.41\\\hline
\end{tabular}
\end{center}
\caption{Comparison of results obtained by counting GP residuals after fitting from 1400 MHz to 1730 MHz obtained by VLBI and using a single dish (Kashima 34 m and Usuda 64 m). Missing values indicate that we could not estimate the DM within 56.75 to 56.77.  The type indicates the main pulse and inter pulse of the Crab pulsar, and the SNR shows a value dedispersed for the first channel of Kashima 34 m with the frequency range of 1400 to 1432 MHz. Most of the VLBI results were about 1.6 times higher than the single-dish observation results.}
\label{tbl:count}
\end{table}

%%%%%%%%%%%%%%%%%%%%%%%%%%%%%%%%%%%%%%%%%%%%%%%%%%%%%%%%%%%%%%%%%%%%%%%%%
%%     FIGURES     %%
\clearpage

\centering

%
%\begin{figure}[hbt]
%  \includegraphics[width=\linewidth]{f5.eps}
%  \caption{Residual components of the GP from the Crab pulsar shown in Figure \ref{fig:grp-bws} after iterative fitting. We count the GP components in the residuals to evaluate with the strength of the GP. }
%  \label{fig:residual}
%\end{figure}

%%%%%%%%%%%%%%%%%%%%%%%%%%%%%%%%%%%%%%%%%%%%

\end{document}